\documentclass[12pt]{article}

\usepackage{epsfig}

\newcommand{\be}{\begin{eqnarray}}
\newcommand{\ee}{\end{eqnarray}}

\def\lsim{\mathrel{\rlap{\lower4pt\hbox{\hskip1pt$\sim$}}
    \raise1pt\hbox{$<$}}}               
\def\gsim{\mathrel{\rlap{\lower4pt\hbox{\hskip1pt$\sim$}}
    \raise1pt\hbox{$>$}}}               

\def\NPA{{\em Nucl. Phys.} A~}
\def\NPB{{\em Nucl. Phys.} B~}

\def\PRC{{\em Phys. Rev.} C~}
\def\PRD{{\em Phys. Rev.} D~}

\begin{document}

\rightline{\Large Preprint HD-THEP-02-35}
\rightline{\Large RM3-TH/02-18}

\vspace{1.5cm}

\begin{center}

\LARGE{End-point singularities of Feynman graphs\\ on the light cone\footnote{{\bf To appear in Physics Letters B.}}}

\vspace{1.5cm}

\large{D. Melikhov$^*$ and S. Simula$^{**}$}

\vspace{0.5cm}

\normalsize{\it $^*$ITP, Universit\"at Heidelberg, Philosophenweg 16, D-69120 Heidelberg, Germany\\[2mm] $^{**}$INFN, Sezione di Roma III, Via della Vasca Navale 84, I-00146, Roma, Italy}

\end{center}

\vspace{1.5cm}

\begin{abstract}

\indent We show that some Lorentz components of the Feynman integrals calculated in terms of the light-cone variables may contain end-point singularities which originate from the contribution of the big-circle integral in the complex $k_-$-plane. These singularities appear in various types of diagrams (two-point functions, three-point functions, etc) and provide the covariance of the Feynman integrals on the light-cone. We propose a procedure for calculating Feynman integrals which guarantees that the end-point singularities do not appear in the light-cone representations of the invariant amplitudes. 

\end{abstract}

\vspace{1.5cm}


\newpage

\pagestyle{plain}

\section{Introduction}

\indent Light-cone representations of Feynman diagrams are widely used in quantum field theory and for the description of $QCD$ bound states \cite{QFT}. It was discussed recently \cite{zero-mode} that the representations of three-point functions in terms of the light-cone variables may contain singular terms of the form $\delta(x)$, $\delta(1-x)$ and their derivatives, sometimes referred to as {\em longitudinal zero modes}. Here $x$ is the fraction of the external light-cone momentum $p_+$ carried by the particle in the loop. 

\indent In this letter we show that end-point singularities may appear in Feynman integrals corresponding to various types of diagrams (two-point functions, three-point functions, etc) calculated using the light-cone variables. The origin of these singularities is related to the integral over the big circle in the complex $k_-$ plane. We study which components of the tensor Feynman integrals may contain end-point singularities and propose a procedure for calculating the invariant amplitudes which guarantees that end-point singularities do not appear explicitly in the integral representations in terms of the light-cone variables. 

\section{End-point singularities in the two-point function}

\indent Let us start with the light-cone calculation of the Feynman integral corresponding to the two-point diagram: 
 \be
    \Sigma_{\mu}(P) & = & \frac{1}{(2\pi)^4 i} \int d^4k ~ 
    \frac{k_{\mu}}{(m^2 - k^2 - i0) ~ [m^2 - (P - k)^2 -i0]^2} 
    \nonumber \\[2mm]
    & = & P_{\mu} ~ f(P^2) ~,
    \label{eq:sigma}
 \ee
where we consider equal masses in the propagators for sake of simplicity. It is convenient to choose the reference frame in which $\vec{P}_{\perp} = 0$ and introduce standard light-cone variables \cite{Dima} for the 4-vector $k = (k_+, k_-, \vec{k}_{\perp})$ with $k_{\pm} = (k_0 \pm k_3) / \sqrt{2}$, so that $k^2 =2 k_ + k_- - k_\perp^2$. The integral (\ref{eq:sigma}) takes the form 
 \be
     \Sigma_{\mu}(P) = \frac{-\pi}{(2\pi)^4i} \int \frac{dk_+ dk_- 
     dk^2_{\perp} ~~~~ k_\mu}{2k_+ \left( k_{-} - \frac{m^2 + k_{\perp}^2 - 
     i0}{2k_+} \right) 2(P_+ - k_+)^2 \left[ P_- - k_- - \frac{m^2 + 
     k_{\perp}^2 - i0}{2(P_+ - k_+)} \right]^2} ~ . 
     \label{eq:sigmamu}
 \ee
Let us write this expression as
 \be
    \Sigma_{\mu}(P) = \int dx ~ dk^2_{\perp} ~ \sigma_{\mu} ~ , 
    \label{eq:sig}
 \ee
and calculate the quantity $\sigma_{\mu}$ by performing the $k_-$-integration. 

\indent The location of singularities in the complex $k_-$ plane is shown in Fig.~1. It is convenient to represent the usual Feynman integral as integral over two contours: one contour $C_R^{Spect}$ encloses the first-order pole (let us call it {\em spectator pole} although there is no real spectator and real interacting particle in the two-point function), while the other integral extends over the contour $C_R^-$ as shown in Fig.~1. The Feynman integral (\ref{eq:sigma}) is obtained by taking in Eq.~(\ref{eq:sig}) the limit $R \to \infty$ with 
 \be
    \sigma_{\mu} = \sigma_{\mu}^{Spect} + \sigma_{\mu}^{Circle} ~ . 
    \label{eq:spect+circle}
 \ee

\begin{figure}[t]

\vspace{-1.5cm}
 
\centerline{\epsfbox{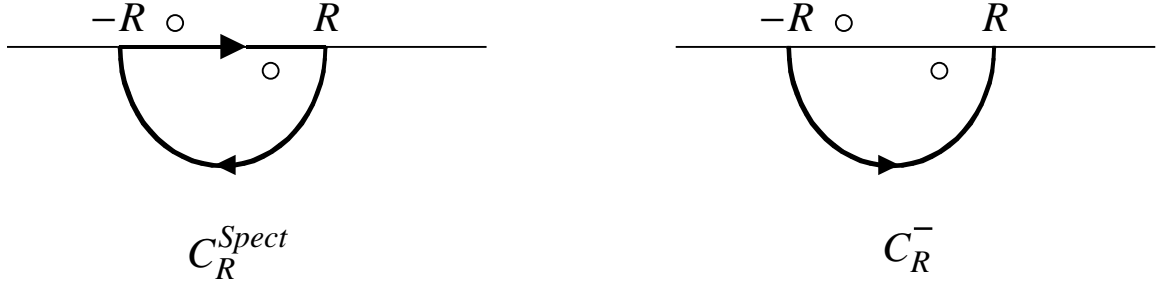}}

\vspace{-1.0cm}

\caption{Location of singularities in the complex $k_-$ plane for the two-point function (\protect\ref{eq:sigma}).}

\vspace{0.25cm}

\end{figure}

\noindent Let us now calculate the spectator and the big-circle contributions, separately. 

\subsection{Spectator pole}

\indent The spectator pole (a pole of the first order in our example) 
gives the following contribution 
 \be
    \sigma^{Spect}_{\mu}(P) & = & \frac{-\pi}{(2\pi)^4i} \int 
    \limits_{C_R^{Spect}} dk_- \frac{k_\mu}{2k_+ \left(k_- - \frac{m^2 + 
    k_{\perp}^2 - i0}{2k_+} \right) 2(P_+ - k_+)^2 \left[ P_- - k_- - 
    \frac{m^2 + k_{\perp}^2 - i0}{2(P_+ - k_+)} \right]^2} \nonumber \\[2mm]
    & = & \frac{1}{16\pi^2} \frac{k_{\mu}^{on}}{P_+} \frac{\theta(x) ~ 
    \theta(1 - x)} {x (1 - x)^2 \left(\frac{m^2 + k_{\perp}^2}{x (1 - x)} - 
    M^2 \right)^2} ~ ,
    \label{eq:spectator}
\ee
where $k_\mu^{on} = (k_+, \frac{m^2 + k_{\perp}^2}{2k_+}, \vec{k}_{\perp})$ is the on-shell spectator 4-vector [$(k^{on})^2 = m^2$]. 

\subsection{Contribution of the big circle}

\indent For $\mu = (+)$ the contribution of the big circle in Fig.~1 vanishes in the limit $R \to \infty$ except for the end-point $x = 1$\footnote{For $\mu = \perp$ the two-point function (\ref{eq:sigma}) is identically vanishing for $\vec{P}_{\perp} = 0$.}. For $x = 1$ the integral over the big circle remains non-vanishing and finite as $R \to \infty$, but this does not lead to any contribution to $\Sigma_{\mu}$. The situation changes however when $\mu = (-)$. Now we get 
 \be
    \sigma_{\mu = (-)}^{Circle}(P) & = & \frac{-\pi}{(2\pi)^4i} \frac{1}{2x} 
    \frac{1}{4P_+^2} \int \limits_{C_R^-} dk_- \frac{1}{(1 - x)^2 \left( P_- 
    - k_- - \frac{m^2 + k_{\perp}^2 - i0}{2(P_+ - k_+)} \right)^2} 
    \nonumber \\[2mm]
    & = & \frac{-\pi}{(2\pi)^4i} \frac{1}{2x} \frac{1}{4P_+^2} \int 
    \limits_{C_R^-} dk_- \frac{1}{\left[ (1 - x) k_- - b \right]^2} ~ ,
    \label{eq:circle}
 \ee
where 
 \be
    b = (1 - x) P_{-}-\frac{m_{\perp}^2 + k_{\perp}^2 - i0}{2P_+} ~. 
    \label{eq:b}
 \ee

\noindent The integral over $k_-$ reads 
 \be
    \int \limits_{C_R^-} dk_- \frac{1}{\left[(1 - x)k_- - b \right]^2} = 
    \frac{-2R}{(1 - x)^2 ~ R^2 - b^2} ~.
    \label{eq:int_k-}
\ee

\noindent Taking the limit $R\to \infty$ we obtain 
 \be
    {\rm lim}_{R \to \infty} ~ \frac{-2R}{(1 - x)^2 ~ R^2 - b^2} = 
    \frac{2\pi i}{b} \delta(1 - x) ~ . 
    \label{eq:lim}
 \ee

\noindent Therefore, the final result takes the form 
 \be
    \sigma_{\mu}^{Spect} & = & \frac{1}{16\pi^2} \frac{k_{\mu}^{on}}{x (1 - 
    x)^2} \frac{\theta(x) ~ \theta(1 - x)}{\left( \frac{m^2 + k_{\perp}^2}{x 
    (1 - x)} - M^2\right)^2} ~ , \nonumber\\
    \sigma_{\mu}^{Circle} & = & - \frac{1}{16\pi^2} \delta_{\mu, -} 
    \frac{\delta(1 - x)}{m^2 + k_{\perp}^2} \frac{1}{2P_+} ~ , 
    \label{eq:result}
 \ee
and $\Sigma_{\mu}$ is obtained by integrating the sum of these expressions over $k_{\perp}$ and $x$. Notice that the $k_{\perp}$-integrals for $\Sigma_{\mu = (-)}^{Spect}$ and $\Sigma_{\mu = (-)}^{Circle}$ logarithmically diverge at large $k_{\perp}^2$, and only their sum gives a finite function $\Sigma_{\mu = (-)}$. 

\indent The invariant amplitude $f(P^2)$ in Eq. (\ref{eq:sigma}) can be calculated from the $(+)$ component, leading to 
 \be
    f(P^2) & = & \frac{1}{16\pi^2} \int \limits_{0}^{1} dx ~ dk_{\perp}^2 ~ 
    \frac{1}{(1 - x)^2 ~ \left[ \frac{m^2 + k_{\perp}^2}{x (1 - x)} - M^2 
    \right]^2} ~. 
    \label{eq:fP_plus}
 \ee

\noindent Using the $(-)$ component one obtains a {\em different integral representation} for $f(P^2)$, but one can check that it still leads to the same function $f(P^2)$ as required by Lorentz invariance. 

\indent It is easy to understand that qualitatively the same happens when one uses light-cone variables for calculating the integrals of the type 
 \be
    \Sigma_{\mu_1 \mu_2 \ldots \mu_m}(P) & = & \frac{1}{(2\pi)^4i} \int d^4k 
    ~ \frac{k_{\mu_1} k_{\mu_2} \ldots k_{\mu_m}}{(m^2 - k^2 - i0)^{n_1}(m^2 
    - (P - k)^2 - i0)^{n_2}} \nonumber \\[2mm]
    & = & \sum_i L_{\mu_1 \mu_2 \ldots \mu_m}^i(P) ~ f_i(P^2) ~,
    \label{eq:sigma_tensor}
 \ee
where $L^i_{\mu_1 \mu_2 \ldots \mu_m}(P)$ are the possible Lorentz tensors and $f_i$ the corresponding invariant amplitudes. Nonzero end-point contributions proportional to $\delta(x)$, $\delta(1 - x)$ and their derivatives may emerge in specific components of the tensor $\Sigma_{\mu_1 \mu_2 \ldots \mu_m}$ if at one of the end-points, $x = 0$ or $x = 1$, the corresponding $k_-$-integral diverges at least linearly. In particular, linear divergence of the $k_-$ integral at $x = 0$ leads to the appearance of the contact term $\delta(x)$, whereas a power divergence of order $s$ leads to the term $\delta^{(s)}(x)$. From the simple example considered above we learn that:  

\begin{itemize}

\item{End-point singularities $\delta(x)$ and $\delta(1 - x)$ in the fraction $x$ of the longitudinal momentum never appear in the components of the tensor $\Sigma_{\mu_1 \mu_2 \ldots \mu_m}(P)$ if {\em all indices are different from $(-)$}.} 

\item{End-point singularities may emerge in the light-cone calculation of the Feynman integrals of the type $\Sigma_{\mu_1 \mu_2 \ldots \mu_m}(P)$ if {\em some of the indices $\mu_i$ are equal to $(-)$}. These end-point singularities emerge as the residual contribution of the big-circle integration in the complex $k_-$ plane.

\noindent One remark is in order here. An alternative way to proceed is as follows. First we modify the denominator of Eq.~(\ref{eq:sigma}):
 \be
    \left[ m^2 - (P - k)^2 \right]^2 \to \left[ m^2 - (P + q - k)^2 
    \right] ~ \left[ m^2 - (P - k)^2 \right] ~,
    \label{eq:displacement}
 \ee
where $q = (q_+, 0, \vec{0}_{\perp})$, and take the limit $q_+ \to 0$ at the final stage of the calculation. In this case one has three poles in the complex $k_-$ plane at different locations, and the calculation looks similar to the triangle diagram for $q_+\ne 0$. In this case the contribution of the big circle is zero, but the equivalent additional 
contribution to $\Sigma_{\mu = (-)}$ re-appears as the nonvanishing residual contribution of the {\em non-spectator pole} in the limit $q_+ \to 0$.}  

\item{End-point contributions provide the covariance of the Feynman two-point integral on the light cone, i.e. they guarantee the independence of the invariant amplitudes $f_i(P^2)$ from the specific components used for their calculation.}

\item{There is an attractive possibility to avoid at all the explicit appearance of end-point singularities in the invariant amplitudes. Namely, as one can check, the number of the independent invariant amplitudes of any tensor $\Sigma_{\mu_1 \mu_2 \ldots \mu_m}(P)$ is precisely equal to the number of non-vanishing $(+)$ and $(\perp)$ components of $\Sigma_{\mu_1 \mu_2 \ldots \mu_m}(P)$.

\noindent Therefore, {\em it is possible to fully extract all the invariant amplitudes by considering only those components which do not involve end-point singularities}. We define such components as the {\em good} components\footnote{Let us point out that our definition of good components differs from the one adopted in Ref.~\cite{FFS}.}.

\noindent For example, in case of $\Sigma_{\mu}(P)$ there is only one form factor which can be determined from $\Sigma_+(P)$; in case of $\Sigma_{\mu_1 \mu_2}(P)$ the two form factors can be determined from $\Sigma_{++}(P)$ and $\Sigma_{\perp \perp}(P)$, etc.}

\end{itemize}

\section{End-point singularities in the three-point function}

\indent Let us now demonstrate that qualitatively the same results hold as well for Feynman integrals corresponding to three-point diagrams. Consider a convergent integral of the following general form
 \be
    \Gamma_{\mu_1 \ldots \mu_n}(p, p'|q) = \int d^4 k ~ \frac{k_{\mu_1} 
    \ldots k_{\mu_n}}{(m^2 - k^2 - i0)^{n_{sp}} ~ [m^2 - (p - k)^2 - 
    i0]^{n_1} ~ [m^2 - (p' - k)^2 - i0]^{n_2}} ~ 
    \label{eq:3point}
 \ee
where the external momenta satisfy the relation $p' = p + q$. The tensor $\Gamma_{\mu_1 \ldots \mu_n}(p, p'|q)$ can be parameterized in terms of the Lorentz covariants constructed using the two independent vectors $p$ and $p'$ and in terms of Lorentz-invariant form factors depending on the three scalars $q^2, p^2$, and $p'^2$ as follows
 \be
    \Gamma_{\mu_1 \ldots \mu_n}(p, p'|q) = \sum_i L^i_{\mu_1 \dots 
    \mu_n}(p, p') ~ f^i(p^2, p'^2, q^2) ~. 
    \label{eq:decomposition}
 \ee

\noindent The integral (\ref{eq:3point}) converges if the powers $n_{sp}, n_1$ and $n_2$ satisfy the relation 
 \be
    n + 4 < 2 ~ (n_{sp} + n_1 + n_2) ~. 
    \label{eq:n}
 \ee

\indent Let us consider the values of the invariants $q^2 < 0$ and $p^2, p'^2 < 4m^2$. We introduce the light-cone variables in the usual way and we use the reference frame in which the external momenta have the following components 
 \be
     q_+ = 0, \qquad \vec{p}_{\perp} = 0 ~.
     \label{eq:frame} 
 \ee

\indent Now, the procedure of calculating the integral (\ref{eq:3point}) just repeats the one used for the two-point function (\ref{eq:sigma}). First, we introduce the light-cone variables for the vector $k$ and translate the Feynman imaginary parts in the propagators to the contour integral in the complex $k_-$ plane while considering $k_+$ and $k_\perp$ to be real variables. Second, we modify the integration contour in the complex $k_-$ plane to isolate the spectator pole. The only difference from the two-point case is that now there are three poles \cite{poles}: the spectator pole of order $n_{sp}$ lies on one side of the integration contour, whereas the two poles related to the interacting particle lie on the other side, as shown in Fig.~2\footnote{Clearly, the situation does not change if instead of one spectator pole $(m^2 - k^2)^{n_{sp}}$ we have $n_{sp}$ single poles corresponding to different masses $(\mu_1^2 - k^2) (\mu_2^2 - k^2) \ldots (\mu_{n_{sp}}^2 - k^2)$.}. So the integration contour has precisely the same form as the one considered in Section 2. We again add and subtract the integral along the big circle and obtain the final result of the $k_-$ integration in the form 
 \be
    \Gamma_{\mu_1 \ldots \mu_n} = \Gamma_{\mu_1 \ldots \mu_n}^{Spect} + 
    \Gamma_{\mu_1 \ldots \mu_n}^{Circle} ~. 
    \label{eq:spectator+circle}
 \ee
As in the two-point case, in the limit $R \to \infty$ the integral over the big circle may diverge at the end-points $x = 0$ or $x = 1$ if some of the indices $\mu_1 \ldots \mu_n$ are equal to $(-)$. This signals the appearance of the end-point singularities in the corresponding components of the tensor (\ref{eq:3point}). It is also clear that no end-point singularities appear in the $\Gamma$ components which contain only {\em good} $(+)$ and $(\perp)$ indices.

\begin{figure}[t]

\vspace{-1.5cm}
 
\centerline{\epsfig{file=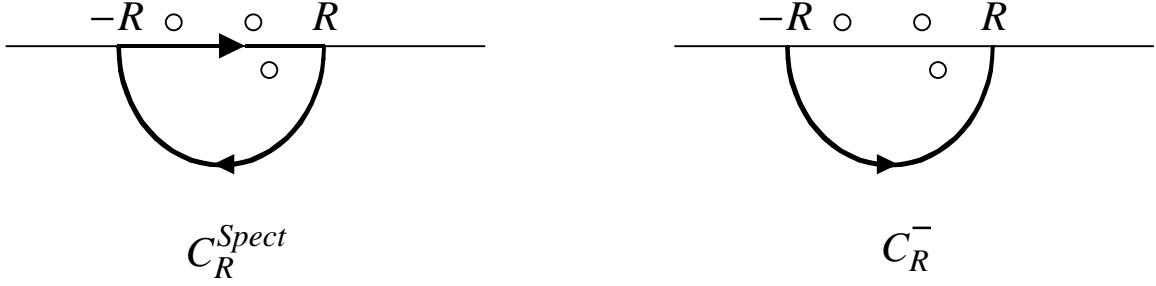}}

\vspace{-1.0cm}

\caption{Location of singularities in the complex $k_-$ plane for the three-point function (\protect\ref{eq:3point}).}

\vspace{0.25cm}

\end{figure}

\indent Let us check that all the $k_{\perp}$ integrals for these {\em good} components are convergent. Potentially the most dangerous case is when all the indices are set $\mu_i = (\perp)$. In this case the integral in $k_\perp$ for large $k_{\perp}$ behaves like 
 \be
    \int dk^2_{\perp} \frac{k_{\perp}^n}{(k_{\perp})^{2 (n_{sp} + n_1 + 
    n_2)}}
    \label{eq:perp}
 \ee
which converges by virtue of Eq.~(\ref{eq:n}). Therefore, both $(+)$ and $(\perp)$ components of the convergent triangle Feynman diagram do not contain any end-point singularities and they can be represented as convergent integrals in $x$ and $k_{\perp}$.

\indent Note that the number of the components of $\Gamma_{\mu_1 \ldots \mu_n}$ containing only $(+)$ and $(\perp)$ indices is equal to the number of the independent form factors. Let us illustrate this statement for small values of $n$: 

\begin{itemize}

\item{$n = 1$. There are 2 form factors which can be determined from the $(+)$ and $(\perp_1)$ components. Hereafter we distinguish between the two components in the transverse plane and denote by $\perp_1$ and $\perp_2$ the components along and perpendicular to $\vec{q}_{\perp}$, respectively. Note that integrals containing an odd number of $\perp_2$ indices are zero.} 

\item{$n = 2$. For a symmetric tensor $\Gamma_{\mu_1 \mu_2}$ there are 4 form factors, which can be found from the $(+~+)$, $(+\perp_1)$ and the two independent $(\perp_1\perp_1)$ and $(\perp_2\perp_2)$ components.}  

\item{$n = 3$. Symmetric tensor $\Gamma_{\mu_1 \mu_2 \mu_3}$ contains 6 independent Lorentz structures, and the corresponding six {\em good} components are $(+~+~+)$, $(+~+\perp_1)$, $(+\perp_1\perp_1)$, $(+\perp_2\perp_2)$, $(\perp_1\perp_1\perp_1)$, $(\perp_1\perp_2\perp_2)$.}

\end{itemize}

\indent Thus, we can always extract all the form factors from the {\em good} components of the Feynman tensor (\ref{eq:3point}) fully avoiding end-point contributions. The latter may appear in the integral representations of the form factors calculated from the $(-)$ components of the three-point function (\ref{eq:3point}). The explicit form of the integral representations depend on which set of components of the tensor $\Gamma_{\mu_1 \ldots \mu_n}$ is used, but of course the form factors $f^i$ are {\em relativistic invariants}, so taht their values do not depend on the specific set of components considered.

\indent We have shown that end-point singularities never appear in the {\em good} $(+)$ and $(\perp)$ components of $\Gamma_{\mu_1 \ldots \mu_n}$, while they may appear in those components which carry the $(-)$ indices. We can easily find under which conditions end-point singularities do not appear in any of the components of the tensor $\Gamma_{\mu_1 \ldots \mu_n}$. The worst situation in the integral (\ref{eq:3point}) occurs if all the indices $\mu_i$ are equal to $(-)$. The corresponding form of the $k_-$ integral is 
 \be
    \int dk_- ~ \frac{k_-^n}{\left[ x k_- \right]^{n_{sp}} \left[ (1 - x) 
    k_- \right]^{n_1 + n_2}} ~.
    \label{eq:k-integral}
 \ee
If
 \be
    n + 1 \le n_{sp} \qquad \mbox{and} \qquad n + 1 \le n_1 + n_2 ~,
    \label{eq:nozeromodes}
 \ee
then for both $x \to 0$ and $x \to 1$ the integral over the big circle remains convergent and no end-point contributions appear at all. 

\indent Before closing this section we point out that in Ref.~\cite{zero-mode} end-point singularities were obtained by means of a different procedure. The Feynman integral (\ref{eq:3point}) was calculated for $q_+ \ne 0$ and the form factors were reconstructed from the $(-)$ components of the tensor. In this case the contribution of the big circle vanishes, but in the limit $q_+ \to 0$ additional contributions to the form factors were found from the non-spectator poles. The longitudinal zero modes obtained in this way were interpreted as a contribution of the pair creation process, prompting that the one-body current approximation is inconsistent. We do not agree with such an interpretation: end-point singularities are not intrinsically related to the pair creation process, because they emerge directly at $q_+ = 0$.

\section{Conclusions}

\indent Let us summarize our main results:

\begin{itemize}

\item[1.]{End-point singularities $\delta(x)$, $\delta(1 - x)$ and their derivatives may appear in the calculations of ultra-violet convergent Feynman integrals using light-cone variables as a remnant contribution of the big-circle integration in the $k_-$-plane. End-point singularities may appear in the components which have some of the indices equal to $(-)$, but they never appear when all the indices are equal to $(+)$ or $(\perp)$. Therefore we define the latter components as the 'good' components.}

\item[2.]{One can avoid the explicit appearance of end-point contributions in the integral representations of the invariant amplitudes in terms of light-cone variables. To this end one should use only the {\em good} $(+)$ and $(\perp)$ components of the Lorentz tensors for extracting the invariant amplitudes. This is always possible since the number of the {\em good} components is equal to the number of the invariant amplitudes. We have followed this strategy in Ref.~\cite{our} for the description of the electromagnetic form factors for spin-0, spin-1/2 and spin-1 bound states.

\noindent One can of course consider $(-)$ components and include the contributions of end-point singularities; then one recovers the same form factors as those obtained from (+) and $(\perp)$ components.}

\item[3.]{End-point singularities are not related to any {\em physical} process. In particular, the zero-mode contribution to the $(-)$ component of the electromagnetic current operator is not related to the pair creation process, although {\em mathematically} the regularizing procedure $q_+\to 0$ is one of the ways to obtain the zero-mode contribution.}

\item[4.]{End-point singularities do not appear in any components of the Feynman integrals if the integrals contain propagators with sufficiently high powers [see Eqs.~(\ref{eq:3point}) and (\ref{eq:nozeromodes})].} 

\end{itemize}

\section*{Acknowledgments} One of us, D.~M., thanks the Alexander von Humboldt-Stiftung and the Istituto Nazionale di Fisica Nucleare for financial support.


\begin{thebibliography}{99}

\bibitem{QFT} S.-J.~Chang and T.-M.~Yan, \PRD {\bf 7}, 1147 (1973). T.-M.~Yan, \PRD {\bf 7}, 1780 (1973). D.~Mustaki, S.~Pinski,J.~Shigemitsu and K.~Wilson: \PRD {\bf 43}, 3411 (1991). M.~Burkardt and A.~Lagnau, \PRD {\bf 44}, 3857 (1991). A.~Langnau and M.~Burkardt, \PRD {\bf 47}, 3452 (1993). For a recent review and further references, see S. J. Brodsky, {{\em Acta Phys. Polon.} B~} {\bf 32}, 4013 (2001) [hep-ph/0111340].  

\bibitem{zero-mode} J.~P.~B.~C.~de Melo, H.~W.~L.~Naus, T.~Frederico, P.~U.~Sauer, \NPA {\bf 660}, 219 (1999) [hep-ph/9908384] and references therein.  

\bibitem{Dima} D.~Melikhov, {{\em Eur. Phys. J. direct} C~} {\bf 2}, 1 (2002) [hep-ph/0110087].

\bibitem{FFS} L. Frankfurt, T. Frederico and M. Strikman, \PRC {\bf 48}, 2182 (1993).

\bibitem{poles} L.~L.~Frankfurt and M.~I.~Strikman, \NPB {\bf 148}, 107 (1979). G.~P.~Lepage and S.~J.~Brodsky, \PRD {\bf 22}, 2157 (1980).
M.~Sawicki, \PRD {\bf 46}, 474 (1992). T.~Frederico and G.~A.~Miller, \PRD {\bf 45}, 4207 (1992) 4207. V~.V.~Anisovich {\em et al.}, \NPA {\bf 563}, 549 (1993). N.~B.~Demchuk {\em et al.}, {\em Phys. of Atom. Nuclei} {\bf 59}, 2152 (1996) [hep-ph/9601369].

\bibitem{our} D.~Melikhov, S.~Simula, \PRD {\bf 65}, 094043 (2002) [hep-ph/0112044]. F.~Cardarelli and S.~Simula, \PRC {\bf 62}, 065201 (2000) [nucl-th/0006023]. S.~Simula, \PRC {\bf 66}, 035201 (2002) [nucl-th/0204015].

\end{thebibliography}
\end{document}